# Pseudospectral continuation for aeroelastic stability analysis

**Arion Pons**
Benin School of Computer Science and Engineering,
and the Silberman Institute of Life Sciences,
Hebrew University of Jerusalem, Giv'at Ram, Jerusalem 91904, Israel
arion.pons@mail.huji.ac.il

**Abstract**

*This technical note is concerned with aeroelastic flutter problems: the analysis of aeroelastic systems undergoing airspeed-dependent dynamic instability. Existing continuation methods for parametric stability analysis are based on marching along an airspeed parameter until the flutter point is found – and approach which may waste computational effort on low-airspeed system behavior, before a flutter point is located and characterized. Here, we describe a pseudospectral continuation approach which instead marches outwards from the system's flutter points, from points of instability to points of increasing damping, allowing efficient characterization of the subcritical and supercritical behavior of the system. This approach ties together aeroelastic stability analysis and abstract linear algebra, and provides efficient methods for computing practical aeroelastic stability properties – for instance, flight envelopes based on maximum modal damping, and the location of borderline-stable zones.*

## Introduction

In the context of eigenvalue problem analysis, Trefethen [1] introduced the concept of *pseudospectra*: the sets of points for which an eigenvalue problem is nearly satisfied. This is usually formulated in terms of an $\epsilon$-pseudospectra: for a linear operator A, generating an eigenvalue problem $(A - \lambda I)\mathbf{x} = 0$, the $\epsilon$-pseudospectra is the set of $\lambda_\epsilon$ such that there exists $\mathbf{x}_\epsilon$ such that $\|(A - \lambda_\epsilon I)\mathbf{x}_\epsilon\| < \epsilon$. That is, the eigenvalue problem is solved to within a residual of $\epsilon$. Pseudospectra have been similarly defined for matrix polynomials [2]. They have been applied to a wide range of physical problems, including control theory [3] and hydrodynamic stability [4]. They provide important and interesting qualitative information about system stability, and hence are of potential use in aeroelastic stability analysis. A key aspect of aeroelastic stability analysis is the computation and characterization of point of airspeed-dependent dynamic instability (flutter points) [5–7]. This characterization is of significant relevance to aircraft safety, and the definition of aircraft flight envelopes [8]. Pons and Gutschmidt [9–11] studied the application of multiparameter spectral theory to aeroelasticity, and showed how flutter instability problems can be recast as a multiparameter eigenvalue problems (MPEVP). The pseudospectra of MPEVPs have been studied in brief [12], and used to characterize flutter point modal damping gradient (hard, soft, *etc.*) [13]. But there are further aeroelastic analysis processes that are amenable to pseudospectral analysis: here, we detail pseudospectral continuation methods that allow efficient characterization of subcritical and supercritical flutter behavior, and the definition of flight envelopes for a maximum acceptable modal damping. These novel pseudospectral methods not only tie together practical aeroelastic stability analysis with abstract linear algebra, but provide efficient methods for computing practical aeroelastic stability properties, such as flight envelopes.

**Definitions and properties of pseudospectra**

Definitions of the $\epsilon$-pseudospectra are directly applicable to multiparameter problems. For an $n$-parameter problem $A(\boldsymbol{\lambda})\mathbf{x} = \mathbf{0}$, the $\epsilon$-pseudospectrum is $\boldsymbol{\lambda}_\epsilon \subseteq \mathbb{C}^n$ such that there exists a normalized $\mathbf{x}_\epsilon$ such that $\|A(\boldsymbol{\lambda}_\epsilon)\mathbf{x}_\epsilon\| < \epsilon$. Via the min-max theorem [14], these minimal $\|A(\boldsymbol{\lambda})\mathbf{x}\|$ values are equivalent to the minimum singular values of $A(\boldsymbol{\lambda})$. To compute these $\epsilon$-pseudospectra, following Trefethen [15], we can generate a field of minimal singular values over the space of $\boldsymbol{\lambda}$ and compute pseudospectral contours. While this method is convenient, however, it is inefficient. The $\epsilon$ definition of the pseudospectrum is also not the most convenient for aeroelastic analysis. It is more obvious to measure distance from stability not in terms of the eigenproblem residual ($\epsilon$) but in terms of a modal damping parameter. A simple way of doing this to introduce an imaginary component into the structural eigenvalue, which in the multiparameter analysis is normally assumed to be real – *i.e.*, purely oscillatory; with an eigenvalue, $\chi$, defined such that $\hat{\mathbf{x}}(t) = \mathbf{x}\exp(\iota\chi t)$. That is, where ordinarily $\chi = \chi_R \in \mathbb{R}$, we take $\chi = \chi_R + \iota\chi_I$, $\chi_R, \chi_I \in \mathbb{R}$. For an arbitrary nonlinear problem $A(\chi, U)\mathbf{x} = \mathbf{0}$, with airspeed parameter $U$, we have simply $A(\chi_R + \iota\chi_I, U)\mathbf{x} = \mathbf{0}$. Note that it follows from Theorem 4.2 of van Dorsselaer et al. [16] that, for a linear problem, $(A - \chi I)\mathbf{x} = 0$, the $\chi_I$-pseudospectrum is equivalent to the $\epsilon$-pseudospectrum: but this equivalence may break down in nonlinear problems.

From the modal damping metric, $\chi_I$, we can define more useful dimensionless parameters, such as the modal damping ratio: $\zeta = \chi_I/\|\chi\|$. The modal damping ratio, $\zeta$, can be substituted directly into a problem, resulting in a system of the form $A(\chi_R, \zeta, U)\mathbf{x} = \mathbf{0}$, but this may introduce nonlinearity into an otherwise linear MPEVP: something which is not significant if $\zeta$ is assumed *a priori*, but may make solving for $\zeta$ more difficult. As a *via media*, we can define the damping parameter $\chi = \chi_R(1 \pm \xi\iota)$, which is both linear and dimensionless; scaling well with the modal frequency ($\chi_R$). Table 1 presents an overview of these three physical pseudospectral parameters. This concept of a flutter pseudospectra is easily generalized: for example, a $k$- or $p$-$k$ method formulation yields naturally the a pseudospectrum in $g$; the artificial structural damping term introduced into the $k$- and $p$-$k$ method systems [17].

Table 1: Modal pseudospectral parameters

| Parameter | Definition | Comments |
| --- | --- | --- |
| $\chi_I$ | $\chi = \chi_R + \iota\chi_I$ | Simple but dimensional, equivalent to $\epsilon$ parameter in some cases. |
| $\zeta$ | $\chi = \chi_R \left(\iota + \frac{\zeta}{\sqrt{1-\zeta^2}}\right)$ | Dimensionless and physically relevant, but introduces strong nonlinearity into the system. |
| $\xi$ | $\chi_R(1 + \xi\iota)$ | Simple and dimensionless. |

**Pseudospectral continuation**

The pseudospectra that we have just defined provide a method of generating the modal damping via numerical continuation along the pseudospectral parameter. At the most basic level this involves specifying a grid of modal damping values and then solving the pseudospectral problem at each of these points. If the system is linear or polynomial this can be done directly [9,13]; if the system is nonlinear then the previous point can supply an initial guess for the iterative solution of the next point. Continuation of this form has been applied to flutter problems before [18–20]: indeed, the traditional modal damping plot corresponds to natural parameter continuation, with the continuation parameter being $U$ or another airspeed parameter, and the pseudospectral parameter (modal damping) being solved for in the eigenvalue problem. However, rather than starting the continuation at zero airspeed and then marching to a flutter point, we can simply compute the flutter points initially via the multiparameter approach [9–11,13], and then expand the modal damping paths out from these points. This produces an accurate picture of the system's supercritical and subcritical behavior, without spending computational effort on areas which not of immediate interest. Continuation with a pseudospectral parameter (*e.g.*, the modal damping ratio) is even more useful as a form of local analysis, as it allows us to specify a modal damping range of interest directly and thus efficiently compute damping-related flight envelopes. However, it also restricts the continuation to areas of monotonic damping behavior: the method will not pass a modal damping turning point.

To overcome this obstacle, higher-order continuation methods are available, such as pseudo-arclength (or Riks') continuation [18,19,21]. Pseudo-arclength continuation is a predictor-corrector procedure; the predictor step involves incrementing $[U, \chi_R, \chi_I]$ along a specified arclength $\Delta s$, according to:

$$\begin{bmatrix} U \\ \chi_R \\ \chi_I \end{bmatrix} = \begin{bmatrix} U_0 \\ \chi_{R,0} \\ \chi_{I,0} \end{bmatrix} + \Delta s \begin{bmatrix} \dot{U}_0 \\ \dot{\chi}_{R,0} \\ \dot{\chi}_{I,0} \end{bmatrix}, \quad (1)$$

using an initial guess $[U_0, \chi_{R,0}, \chi_{I,0}]$ and the local tangent $[\dot{U}_0, \dot{\chi}_{R,0}, \dot{\chi}_{I,0}]$. We compute this tangent using a finite-difference approximation based on the previous point computed on the continuation curve. The corrector step involves solving for the system's local modal properties subject to a pseudo-arclength constraint. However, instead of the larger system of nonlinear equations employed Meyer [18,19], we can use the pseudospectral approach to formulate the corrector step as three-parameter eigenvalue problem:

$$\begin{aligned} A(\chi_R, \chi_I, U)\mathbf{x} &= \mathbf{0}, \\ \bar{A}(\chi_R, \chi_I, U)\bar{\mathbf{x}} &= \mathbf{0}, \\ (\dot{U}_0(U - U_0) + \dot{\chi}_{R,0}(\chi_R - \chi_{R,0}) + \dot{\chi}_{I,0}(\chi_I - \chi_{I,0}) - \Delta s)y &= 0, \end{aligned} \quad (2)$$

where the conjugate equation enforces $\chi_R, \chi_I \in \mathbb{R}$, as per [9,10], and the pseudo-arclength constraint [21] has been recast as an MPEVP using an arbitrary scalar eigenvalue $y$. This is a three-parameter nonlinear eigenvalue problem which we solve with the method of successive linear problems [10]. We note in passing that, for the SLP algorithm, a significantly simpler form of the constraint in Eq. 2 is:

$$(\dot{U}_0 \Delta U + \dot{\chi}_{R,0} \Delta \chi_R + \dot{\chi}_{I,0} \Delta \chi_I)y = 0, \quad (3)$$

with $\Delta x = x_{k+1} - x_k$ as per [22]. This continuation approach can easily be generalized to other sets of parameters, for instance wing and airframe structural parameters.

**Application**

For demonstration, we analyze the well-established Goland wing test case, with model parameter values taken from [11]. We reduce the Goland wing partial-differential model to an algebraic MPEVP not via discretization, but via the Generalized Laplace Transform Method [11]: the result is an unstructured nonlinear MPEVP in the modal frequency ($\chi$) and airspeed ($U$). The problem is then restricted [22] to simplify the system behavior at nonphysical parameter values ($\chi < 0$ and $U < 0$). We compute the first flutter point of this system using the iterated contour plot algorithm [10], and then, using the $\epsilon$-pseudospectrum, compute the set of points with $\epsilon = 0.04$ and $\epsilon = 0.08$. Figure 1 illustrates these results, visualized over a contour plot of the system [11]: the intersections of real and imaginary contours represent flutter points. One particularly interesting feature is observed: a pseudospectral area (*i.e.*, a mode near the stability boundary) at $U = 600$ m/s, not associated with any flutter point. The $\epsilon$-pseudospectrum is a powerful tool to locate these borderline-stable areas, whether nearly-stable or nearly-unstable. With modal pseudospectra, we can characterize their stability in more detail.

To do this, we apply our modal continuation method. We then compute the mode paths of the system via pseudo-arclength continuation, starting from two areas of interest: the first flutter point and the pseudospectral area noted in Figure 1. Short continuation paths are extended from these points: these are shown in Figure 2. Note that the pseudo-arclength method operates in the $\chi_I$ pseudospectral definition: other variables produce poorly-scaled arclength definitions due to the difference in magnitude between $\zeta$ and $\chi_R$. Results in $\chi_I$ are postprocessed into $\zeta$ for Figure 1. Using only one continuation point we can both compute a flight envelope around the flutter point (*e.g.*, for $\zeta > 0.05$, $U < 125.6$ m/s). Further continuation provides information on the wider modal trends: for instance, that the pseudospectral area noted in Figure 1 represents a near-restabilization, reaching $\zeta = -0.08$ maximum. Extending the mode further, or analyzing the modeshape, we observe that the restabilizing mode is the same mode destabilized at the first flutter point. Pseudospectral analysis is particularly useful for locating and characterizing these borderline-stable areas: practically, these borderline stable zones could represent areas in which the flutter instability is easily controlled via an active suppression system [5].

**Conclusion**

In this work we presented a novel pseudospectral continuation method for the analysis of aeroelastic flutter problems. Using a range of different pseudospectral formulations, we described continuation methods for flutter analysis that are able to focus computational effort on the vicinity of flutter points, and demonstrated these methods in the context of the Goland wing benchmark system. These pseudospectral continuation methods tie together aeroelastic stability analysis and abstract linear algebra, and are particularly useful for efficiently computing of advanced secondary flutter results, such as flight envelopes based on maximum modal damping, flutter onset behavior, and the location and nature of borderline-stable states.

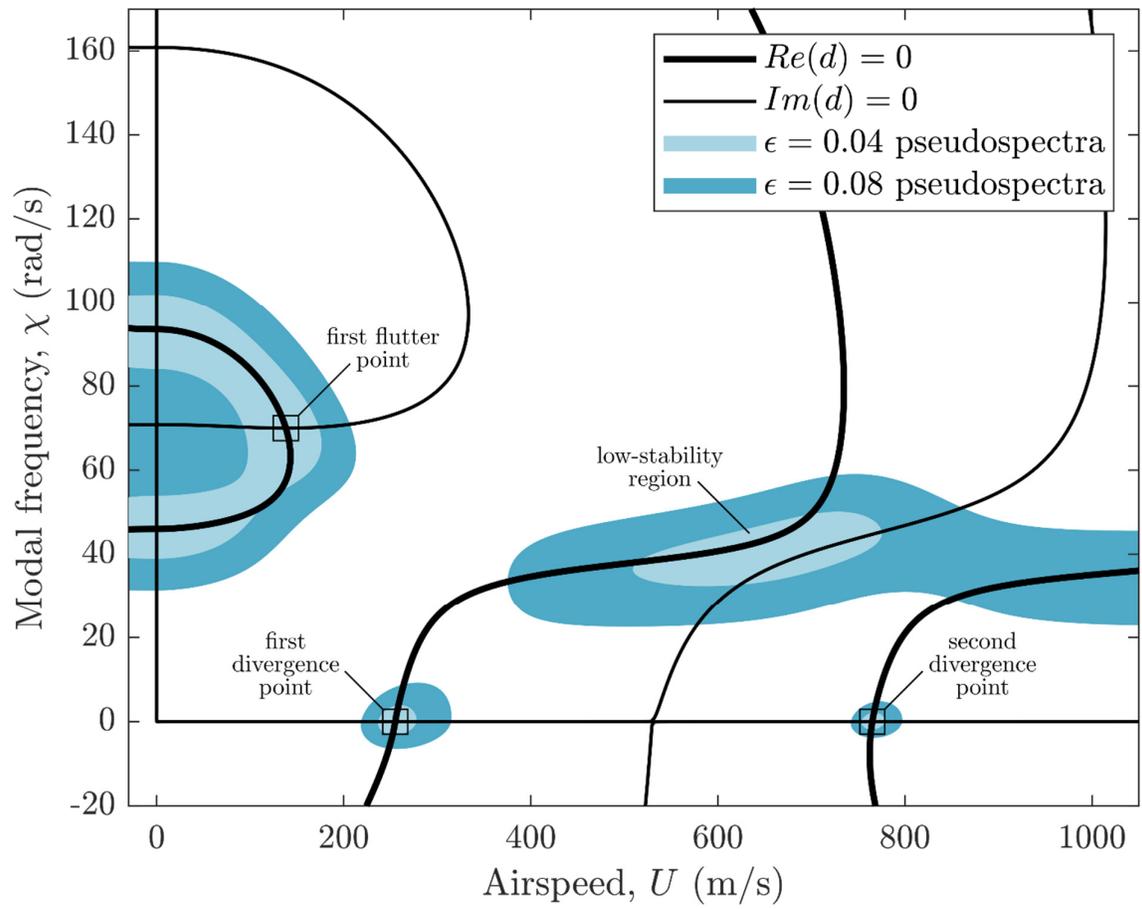

**Figure 1**

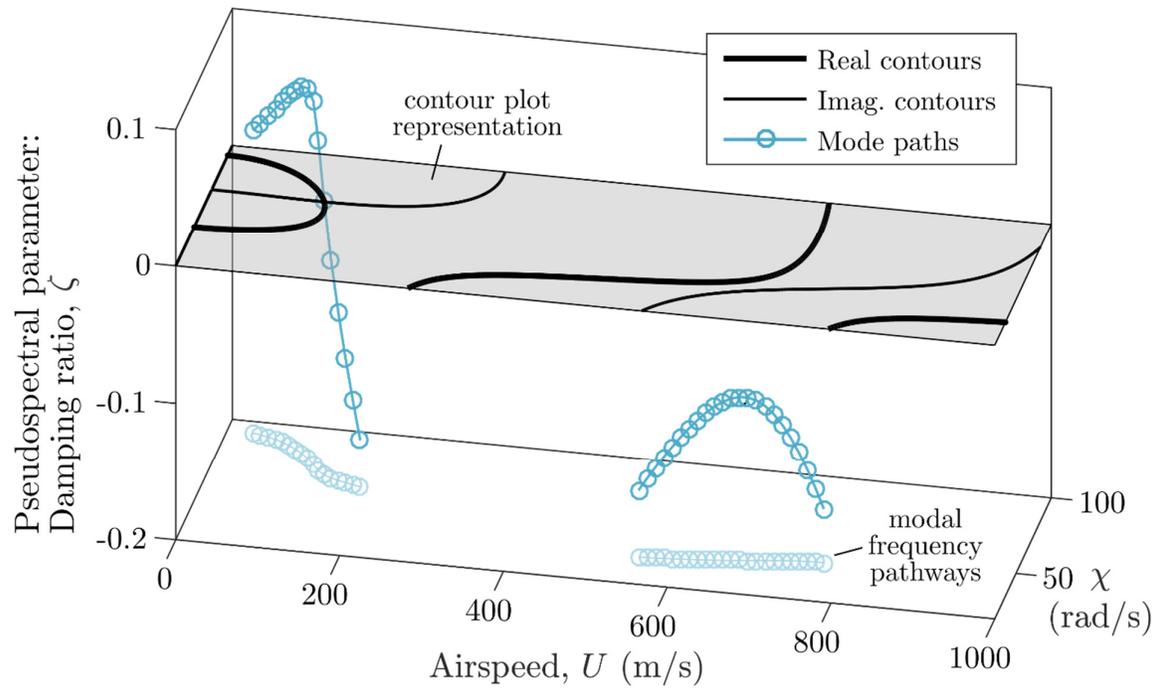

**Figure 2**

**Figure Captions List**

Fig. 1      $\epsilon$-pseudospectra of the Goland wing model, $\epsilon \leq 0.04$, and $\epsilon \leq 0.08$, illustrating the exact instability points: both flutter (dynamic instability) and divergence (static instability). The $\epsilon$-pseudospectra also provide information on the location of regions of low stability away from these instability points – as indicated.

Fig. 2      Modal paths of the Goland wing model in areas of interest, generated via pseudo-arclength continuation. Projections of these paths onto the $U$-$\chi$ plane (modal frequency paths) are illustrated, alongside the contour plots of Fig. 1.

**Table Caption List**

Table 1            Modal pseudospectral parameters